\documentclass[12pt]{article}
\usepackage{cite}
\usepackage{times} \usepackage{amssymb}
\usepackage{pifont}\usepackage{amsmath}
\usepackage{subfigure}
\usepackage{pst-grad}
\usepackage{pst-plot}
\usepackage{pst-slpe}
\usepackage{pst-3dplot}
\usepackage[mathscr]{eucal}
\usepackage[all]{xy}
\xyoption{arc}
\xyoption{curve}
\numberwithin{equation}{section}
\newtheorem{exx}{Example}[section]
\newenvironment{example}{\begin{exx}
\par\normalfont}{\null\hfill\ding{113}\end{exx}}
\def\A{\mathcal{A}}
\def\C{{\mathbf{C}}}
\def\tA{\mathscr{A}}
\def\tB{\mathscr{B}}
\def\tC{\mathscr{C}}
\def\tD{\mathscr{D}}
\def\tE{{\mathscr E}}
\def\tF{\mathscr{F}}
\def\heart{\mbox{\tiny\ding{170}}}
\newcommand\bul[1]{{#1}^{\bullet}}
\newcommand\ch[1]{\operatorname{ch}({#1})}
\newcommand\coh{\operatorname{Coh\,}}
\def\cp#1{{\mathbf{P}}^{#1}}
\newcommand\dg[1]{D^{\geq #1}}
\newcommand\dl[1]{D^{\leq #1}}
\newcommand\dercat[1]{D^{b}(#1)}

\def\eq#1{(\ref{#1})}
\newcommand\figref[1]{Figure~\ref{#1}}

\newcommand\Hom{\operatorname{Hom}}

\def\M{\mathcal{M}}
\def\pa{\partial}
\def\W{{\mathcal W}}
\def\rt{\longrightarrow}
\def\R{{\mathbf{R}}}
\def\T{\mathbf{T}}
\def\td#1{\operatorname{Td}({#1})}
\def\tr{\operatorname{Tr}}
\def\viz{{\slshape viz.~}}
\def\vphi{\varphi}
\def\Z{\mathbf{Z}}
\newfont\sheafnt{rsfs10}
\def\sheaf#1{{\mbox{\sheafnt #1}}}
\renewcommand{\O}{{\sheaf{O}}}

\newcommand\email[4]{#1@#2.#3.#4}
\begin{document}
\title{Branes in hearts with perverse sheaves}
\author{ 
Subir Mukhopadhyay
\thanks{\email{subir}{iopb}{res}{in}} \\
\small Institute of Physics, Bhubaneswar 751~005, India.
\\ 
Koushik Ray 
\thanks{\email{koushik}{iacs}{res}{in}}\\
\small Department of Theoretical Physics\\
\small  Indian Association for
 the Cultivation of Science\\
\small  Calcutta 700 032, India.
}
\date{}
\maketitle
\begin{abstract}
\small
\noindent 
Various topological properties of  
D-branes in the type--IIA 
theory are captured by the  topologically twisted
B-model, treating D-branes as objects in the bounded derived 
category of coherent sheaves on the compact part of the target space. 
The set of basic D-branes wrapped on the homology cycles of 
the compact space are taken to reside in the heart of t-structures of the 
derived category of coherent sheaves on the space
at any point in the K\"ahler moduli space. The stability data
entails specifying a t-structure along with a grade for sorting the
branes. Considering an example of a degenerate Calabi-Yau space, 
obtained via geometric engineering, that retains
but a projective curve as the sole non-compact part, we identify the regions
in the K\"ahler moduli space of the curve that pertain to 
the different t-structures of the bounded derived category 
of coherent sheaves on the curve corresponding to the different phases
of the topological branes.
\end{abstract}
\thispagestyle{empty}
\clearpage
\tableofcontents
\section{A physically inclined overture}
\label{sec:intro}
A class of objects in the spectrum of string theory spatially extended 
beyond a single dimension have received the appellation,
D-branes\cite{Pol1}. These objects have played a pivotal role in 
the exploration of various non-perturbative aspects of string
theory and its sundry consequences
\cite{SenNonP,PCJ,Johnson1,Johnson2,Johnson3}. 
Depending on the type or version of string theory one considers and
the point of view, D-branes are described by different types of data. 
From a conformal field theoretic point of view, for example,
D-branes correspond to Dirichlet boundary conditions imposed on the 
end-points of open strings in a classical picture, lending the suffix 
\emph{D} to the neologism. In a low-energy target-space
description, on the other hand, D-branes are described as certain
extended solitonic solutions of the effective
gravitational theories, carrying the
Ramond-Ramond fields in supersymmetric instances.
A D$p$-brane is an object extending in $p$
space directions, the part \emph{brane} having been extracted from a
two-dimensional \emph{membrane} of which it is imagined to be a 
dimensionally generalized form. A related
description of a D$p$-brane is in terms of the Dirac-Born-Infeld theory on
its $(p+1)$-dimensional world-volume, which in appropriate limits 
of coupling reduces to a gauge theory. The geometric character 
of the gravitational
or gauge theories has instigated endeavors to unearth  
a canonical definition of D-branes in string theory in terms of purely 
geometric data. This has
been achieved for certain restricted classes of D-branes in certain
classes of string theories, no ordinary a feat, considering the
intricacies involved, as we shall remark in the sequel. In this 
article we discuss the definition of BPS 
D-branes in the topological
B-model as certain stable objects in the bounded derived category of coherent
sheaves on the target-space. Before plunging into this, let
us briefly discuss some of the physical properties of D-branes and 
examine the inadequacy of the possible ``intuitive" geometric
definitions within the context of the theories of which allusion has 
been made above. 

\begin{figure}[h]
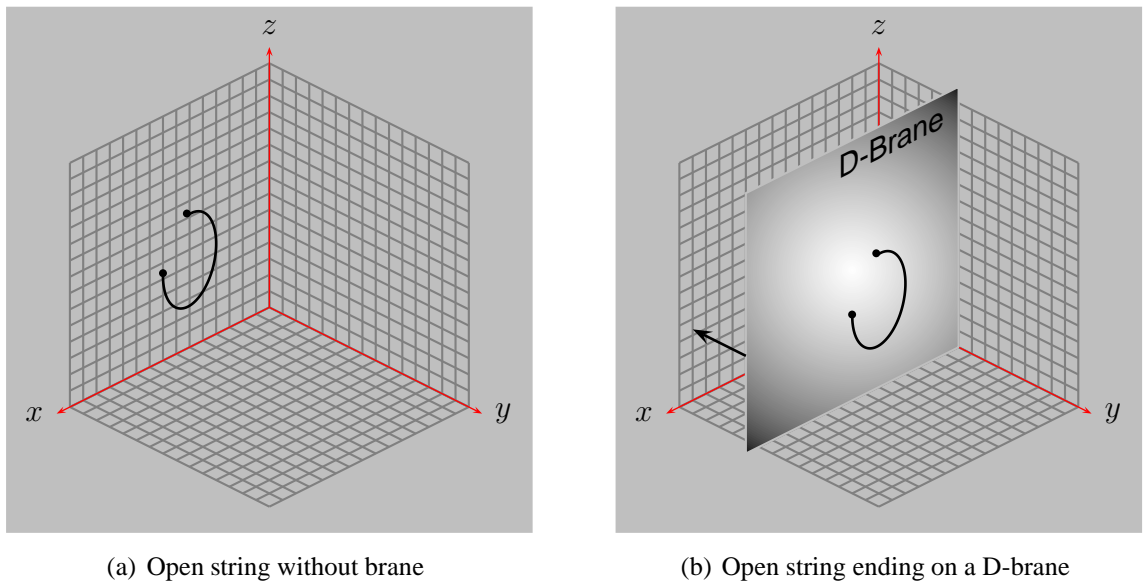

\subfigure[Open string without brane]{%
\label{cartoon1a}
\pspicture(-4,-3)(4,4)
\psset{unit=.5cm}
\psframe*[linecolor=lightgray](-7,-6)(7,8)
\pstThreeDPlaneGrid[subticks=15,linecolor=gray](0,0)(7.5,7.5)
\pstThreeDPlaneGrid[subticks=15,linecolor=gray,planeGrid=xz](0,0)(7.5,7.5)
\pstThreeDPlaneGrid[subticks=15,linecolor=gray,planeGrid=yz](0,0)(7.5,7.5)
\pstThreeDCoor[xMin=0,yMin=0,zMin=0,xMax=8,yMax=8,zMax=8]
\pstThreeDDot[linecolor=black](5,1,3.5)
\pstThreeDDot[linecolor=black](5,1.9,5.7)
\psset{beginAngle=0,endAngle=270,linecolor=black,linewidth=1pt}
\pstThreeDEllipse(4,1,3.5)(-.5,.5,.5)(.5,.5,-1)
\endpspicture
}%
\subfigure[Open string ending on a D-brane]{%
\label{cartoon1b}
\pspicture(-4,-3)(4,4)
\psset{unit=.5cm}
\psframe*[linecolor=lightgray](-7,-6)(7,8)
\pstThreeDPlaneGrid[subticks=15,linecolor=gray](0,0)(7.5,7.5)
\pstThreeDPlaneGrid[subticks=15,linecolor=gray,planeGrid=xz](0,0)(7.5,7.5)
\pstThreeDPlaneGrid[subticks=15,linecolor=gray,planeGrid=yz](0,0)(7.5,7.5)
\pstThreeDCoor[xMin=0,yMin=0,zMin=0,xMax=8,yMax=8,zMax=8]
{\psset{fillstyle=ccslope,slopebegin=white,slopeend=black,slopeangle=45,linecolor=lightgray}
\pstThreeDSquare(0,3,0)(8,0,0)(0,0,8)
}
\pstThreeDDot[linecolor=black](5,4,3.45)
\pstThreeDDot[linecolor=black](5,4.9,5.7)
\psset{beginAngle=0,endAngle=270,linecolor=black,linewidth=1pt}
\pstThreeDEllipse(4,4,3.5)(-.5,.5,.5)(.5,.5,-1)
\pstThreeDLine[arrowscale=1.5,linecolor=black]{->}(8,3,3)(8,1,3)
\pstPlanePut[plane=xz,planecorr=normal](4.5,3,7){\sf{D-Brane}}
\endpspicture
}%
\caption{Open strings with and without D-branes}
\end{figure}

Let us begin with the world-sheet description of D-branes. The
importance of this description lies in its ubiquity, as 
the validity of any geometric definition is to be tested against
expectations from a conformal field theoretic description
\cite{tapo,tj}.
The non-linear sigma-model action describing the world-sheet of a 
string is\cite{gsw12,Polbook}
\begin{equation}
\label{bos}
{\mathcal S} = \int\limits_S d^2\sigma \left(g_{MN}+b_{MN}\right)
\eta^{ij}\pa_iX^M\pa_iX^N,
\end{equation}
where $i,j =0,1$; $\sigma^0$ and $\sigma^1$ denote, respectively, the
temporal and the spatial coordinates of the world-sheet 
of the string and $\eta$ denotes the flat metric on the world-sheet. The
fields $X^M$ are interpreted as the coordinates of the target-space,
with $M,N$ ranging over the dimensions of the target-space, ten for
superstrings and twenty-six for bosonic strings, with a fermionic
piece for the former, which will not feature in our discussion at
this point.
The coupling parameters of the sigma-model constitute a symmetric
matrix $g_{MN}$, interpreted as 
the metric on the target-space and an anti-symmetric one, $b_{MN}$, 
known as the Kalb-Ramond field, giving rise to torsion in the target
space. The integration is over the area of
the world-sheet, $S$. The Euler-Lagrange equations ensuing from this action
are the two-dimensional Laplace's equations one for each field $X^M$,
\begin{equation}
\eta_{ij}\pa_i\pa_j X^M =0.
\end{equation}
For open strings, that is strings extending between
$\sigma^1=0$ and $\sigma^1=\pi$, we can solve the Laplace's equation by
choosing either a Neumann or a Dirichlet boundary condition at the
edges. Moreover, since the fields $X^M$ are independent of each other,
we can choose different boundary conditions for the different $X^M$. 
While choosing the Neumann condition for some $X^M$ leaves the
end-point of the open
string dangling in that direction of the target-space, 
the choice of Dirichlet condition for some of the fields requires
fixing the edge at a certain point in the respective directions,
as shown in \figref{cartoon1a}, 
resulting into a breakdown of Poincar\'e invariance in the
target-space, the \emph{space-time}. 

An awkward predicament as such may be avoided by assuming that
the string is not stuck at a special point in the space-time, but on 
an \emph{object} that perambulates the target-space, as
illustrated in \figref{cartoon1b}. This object is a D-brane. 
If we consider an open string with the end $\sigma^1=0$ on a D-brane,
then the dimension $p$ of the D$p$-brane equals the number of $X$'s on which
Dirichlet condition is imposed at $\sigma^1=0$. Thus, a superstring
theory may have D-branes with maximal dimension nine, which pervades all
of space. In the target-space, then, it appears that a D-brane can
reside in a part of the space-time, a \emph{subspace}. 

Gauge fields provide further embellishment to this picture of D-branes.
An end-point of an open string may support gauge
degrees of freedom, known as Chan-Paton factors. In the world-sheet
description, this is incorporated by adjoining a term 
\cite{gsw12,Polbook,Johnson3}
\begin{equation}
{\mathcal S}_{\mathrm{gauge}} = \tr\int\limits_{\pa S} A_M \pa_0 X^M
\end{equation}
to the action $\mathcal S$ in \eq{bos}. Here $A_M$ denotes a gauge field
and the integration is over the boundary of the world-sheet, $\pa S$,
the trace being over the gauge indices. The corresponding effective
field theory yields the world-volume theory of D-branes as the
Dirac-Born-Infeld action\cite{DLP,Polbook},
\begin{equation}
\label{DBI}
{\mathcal S}_{\mathrm{BI}} = \tr\int\limits_V \sqrt{\det(G+B+F)},
\end{equation}
where $G$ is the metric induced from the space-time on the world-volume
of the brane, $B$ an anti-symmetric field induced from the
Kalb-Ramond field and $F$ denotes the two-form field-strength 
corresponding to $A$.
The integration is over the world-volume of the brane, $V$. The first
non-trivial leading order term in the expansion of the square-root
yields a Yang-Mills theory on the world-volume of the brane. Thus,
supplementing the earlier description, it
appears that a D$p$-brane may be described as a vector bundle on a
$p$-dimensional subspace of the target-space. 

Let us now proceed to consider compactification of string theory and dualities.
From now on we restrict ourselves to superstring
theory only. First, let us consider a simple case, a supersymmetric BPS 
D4-brane in the type--IIA theory compactified on a two-dimensional 
torus, $\T^2$. Let us assume that
the brane lies along the directions of $X^6$, $X^7$, $X^8$ and $X^9$ of
the target-space and the torus is along  $X^4$-$X^5$. If we now perform
two T-duality transformations along the two directions of
the torus seriatim, then since T-duality exchanges Neumann 
with Dirichlet boundary condition, the brane extends
along the torus, turning into a D6-brane. However, T-duality is
assumed to be a symmetry of string theory and two such consecutive operations 
are supposed to leave the type--IIA theory unaltered. Hence the two theories 
must be identified along with their spectra. In other words,
in the type--IIA theory on a torus, the transmogrification of a 
D4-brane into a D6-brane must be allowed. This example 
exhibits that duality transformations do not respect dimensionality of
branes; dimensions of branes change as we change the parameters of
compactification, the \emph{moduli}, of the target-space, perhaps by a
T-duality transformation.
Thus the description of D-branes as (bundles
on) some subspace of \emph{fixed} dimension of the target is inadequate.

Next, let us consider the example of the type-IIA theory on the
target-space $\R^{1,3}\times\M$, where $\M$ is a six-dimensional
Calabi-Yau manifold. Let us consider a BPS D0-brane in this
four-dimensional theory. 
A BPS brane is a supersymmetric solution of the corresponding
supergravity theory obeying the BPS conditions. Many of the 
topological properties of these are
described by the topologically twisted B-model, the branes being B-branes. 
Since the BPS branes in the type--IIA theory are
even-dimensional, and let us note that this specification is free from
the caveat alluded to above, the D0-brane in
four-dimensions may arise from a D6-brane wrapping the six-dimensional
manifold $\M$ or a D$p$-brane wrapping a supersymmetric $p$-cycle of
the homology of $\M$, for $p=4,2,0$. As discussed above, in attempting to
describe the D0-brane as a vector bundle, it can be viewed as
either a vector bundle above $\M$ or one of these cycles or their
combination. However, as we change the size of the Calabi-Yau manifold,
by changing the available K\"ahler structure moduli, cycles in the homology
transmute among each other. For example, a four-cycle may go into a
combination of a four-cycle and a two-cycle or vice versa. Thus, 
bundles on cycles is not a very useful specification; we need to
describe the bundles in terms of structures on 
the six-dimensional Calabi-Yau
manifold itself. This can be achieved by 
extending a bundle on a plucklower dimensional cycle to the
whole of $\M$ by a zero section --- so viewed over $\M$ the rank of the
bundle jumps. We are thus led to consider sheaves on the manifold
$\M$. The collection of branes in such a scenario leads us to
considering a category of sheaves on $\M$, the open string
stretched between branes providing the morphisms in the category.
Transmutation of cycles can be thought of as formation or
decay of bound states of branes, when branes are taken to be these
sheaves. 

In the collection of branes we need to include anti-branes too. A brane and its
corresponding anti-brane differ by a change of sign of the charge they
carry. Considering branes and anti-branes within the same schemata
forces us to consider not only sheaves but complexes of sheaves. We
shall discuss more on this in the next section. 
pluck
Now, as we deform $\M$, there are points in the K\"ahler moduli
space of $\M$ where a brane decays or branes form bound states, as
mentioned above. Right at this point, called the point of \emph{marginal
stability} 
\cite{sen2,mrd:ictp,douglas1,Sharpe:1999qz,douglas2,douglas3,dfr1,dfr2},
the brane or the collection of its
decay products or components can not be distinguished. 
Incorporation of this indistinguishability
necessitates the identification of a sheaf and its resolutions,
ushering in the appearance of derived functors into the arena.
Consistency with the conjectured mirror symmetry, which is envisaged 
as a geometric realization of T-duality, at least in part, 
requires these sheaves to be coherent. 
We are thus led to consider the derived category of coherent sheaves 
\cite{brid4} on $\M$.
In addition, in the physically interesting cases a BPS brane decays into a
finite number of products or bound states of only a finite number of
branes are considered. Tpluckhis
restricts us to consider finite complexes of sheaves. We thus arrive
at the bounded derived category of coherent sheaves on $\M$.
Finally, BPS branes are stable in a certain
sense; unstable configurations decay to a stable one. We shall
expatiate on the stability criterion in the next section. 
In the case of type--IIB theory arguments similar to the 
ones above are valid. But in
that case one needs to consider odd-dimensional cycles and the 
description of the
category of branes is more complicated.

To summarize, we have motivated the description of BPS D-branes in the
type--IIA theory as stable objects in the derived category of 
sheaves on a compact part of the target-space.
We now proceed to discuss the concepts further in the next section.
\section{Viewing homologically}
\centerline{\emph{\ding{125}A derived category is ... when you take 
complexes seriously\ding{126}} \cite{scott}.}
In this section we recall \cite{scott}
some features of derived categories relevant
for our discussion of physical situations 
in the following sections
\cite{scott,manin,hars,brid1,brid2,brid3,brid4,aspin1,
mrd:ictp,douglas1,Sharpe:1999qz,douglas2,douglas3}.

Let $\A$ denote an \emph{Abelian category}, that is, a
category with kernels and co-kernels of
morphisms defined within. Given an Abelian category $\A$, the
corresponding \emph{derived category} $D(\A)$ consists of complexes
$A^{\bullet}$ of objects of $\A$, \viz
\begin{gather}
\xymatrix{
A^{\bullet} = 
&\cdots\ar[r]^{d_{n-2}} & A_{n-1}\ar[r]^{d_{n-1}}& A_n
\ar[r]^{d_n} & A_{n+1}\ar[r]^{d_{n+1}}& \cdots
}\\
 d_n \circ d_{n+1} = 0, \quad A_i\in\A, \nonumber
\end{gather}
upto identification by quasi-isomorphisms. A \emph{quasi-isomorphism}
is a morphism of complexes inducing an isomorphism on cohomology. A
derived category is \emph{bounded} if the complexes have only a finite
number of non-vanishing elements. A bounded derived category
corresponding to $\A$ is denoted $D^b(\A)$.
If $A^{\bullet}$ represents an object in $D^b(\A)$, then it is
quasi-isomorphic to a certain complex $I^{\bullet}$ of injectives
bounded below, 
\begin{equation}
\xymatrix{0\ar[r] & A^{\bullet}\ar[r] & I^{\bullet}}.
\end{equation}
\begin{example}
Let $A^{\bullet}= \xymatrix{\cdots\ar[r]&\underline{A}\ar[r]&\cdots}$
be a complex concentrated in degree zero, the underlined entry. 
A quasi-isomorphism
$A^{\bullet}\rt I^{\bullet}$ is an injective resolution of $A$,
that is, an exact sequence, 
\begin{equation*}
\xymatrix{0\ar[r]&A\ar[r]^{\varepsilon}&
I_0\ar[r]^{d_0}&I_1\ar[r]^{d_1}&\cdots}
\end{equation*}
Then the cohomologies are isomorphic, 
$H^{\star}(I^{\bullet}) \simeq H^{\star}(A)$, where one defines 
$H^0(I^{\bullet})=\ker d_0 \simeq \mathrm{Im}~\varepsilon = H^0(A)$.
\end{example}
Two injective resolutions of $A$  are isomorphic to $A$ in the derived
category and thus to each other, implying that they are homotopy
equivalent. These statements continue to be valid, mutatis mutandis, 
with $A$ replaced by a complex $A^{\bullet}$ in the above example. 
Physically, a D-brane corresponds to a complex $\bul{A}$ in $D^b({\A})$ for
some Abelian category $\A$, to be specified later.
\subsection{Shifts and triangles of complexes}
Among the structures a derived category is endowed with are 
\emph{translations} and \emph{triangles}.  
A translation is a functor defined as a shift in the left of the
entries of a complex, namely, 
\begin{equation}
\begin{split}
[n]: A^{\bullet} &\rt A^{\bullet}[n], \\
A[n]_m &= A_{m+n}.
\end{split}
\end{equation}
\begin{example}
Defining the \emph{mapping cone} of a morphism
$\xymatrix{A^{\bullet}\ar[r]^f& B^{\bullet}}$ between two complexes 
as the complex $M(f) = A^{\bullet}[1]\oplus B^{\bullet}$ with
appropriately shifted morphisms, we naturally
obtain the complex
\begin{equation}
\label{mcone}
\xymatrix{A^{\bullet}\ar[r]^f & B^{\bullet}\ar[r]& M(f)\ar[r] &
A^{\bullet}[1]},
\end{equation}
where the last term is the complex $A^{\bullet}$ shifted by unit
degree. 
\end{example}
Physically, if a D-brane corresponds to a complex $\bul{A}$, then
the shifted complex $\bul{A}[1]$, and any of its cousins shifted by 
an odd degree thereof, corresponds to the anti-D-brane. 
The justification for this interpretation is derived from the fact
that, if we define the Chern character map from $D^b(\A)$ to
cohomology, then $\ch{\bul{A}[n]}=(-1)^n\ch{\bul{A}}$, for an integer
$n$ and the charge of a D-brane $\bul{A}$ is taken to be 
proportional to $\ch{\bul{A}}$.
Moreover, $M(f)$ is taken to represent the marginal
deformation of the configuration consisting of the two 
D-branes corresponding to $\bul{A}$ and $\bul{B}$.

Triangles in the derived category, on the other hand, are counterparts 
of the exact sequences in an Abelian category. 
\begin{example}
Let us consider a short exact sequence $\xymatrix{0\ar[r]& A\ar[r]& B
\ar[r]& C\ar[r]& 0}$ in the Abelian category $\A$. Let us form a
complex $C^{\bullet} =
\xymatrix{\cdots 0\ar[r]&A\ar[r]&\underline{B}\ar[r]&0\cdots}$.
Then this complex is isomorphic to $C$ in the derived category,
$C^{\bullet}\simeq C$, as discussed above. For this realization
$C^{\bullet}$ of $C$ we have the map $C^{\bullet}\rt A[1]$, namely,
\begin{equation*}
\xymatrix{
C^{\bullet}\ar[d]& = 
& \cdots 0 \ar[r]&A\ar[r]\ar[d]_1&\underline{B}\ar[d]_0\ar[r]
&0\cdots\\
A[1] &=&   \cdots 0 \ar[r]&A\ar[r]&\underline{0}\ar[r]&0\cdots}
\end{equation*}
This completes the triangle 
$\xymatrix{A\ar[r]&B\ar[r]&C\ar[r]&A[1]}$, also written in a more
picturesque form as
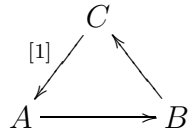
\begin{equation*}
\xymatrix@C5mm{&C\ar[dl]_{[1]}&\\A\ar[rr]&&B\ar[ul]}
\end{equation*}
The occurrence of triangles makes the derived category into a
triangulated category. Again, all the statements above are valid when
the objects are replaced by complexes.
\end{example}
A triangle is called \emph{distinguished} if it is quasi-isomorphic to
\eq{mcone}.
While D-branes are represented by complexes, an oriented open string stretched
between two D-branes $\bul{A}$ and $\bul{B}$ is taken to be
represented by a distinguished triangle.
The triangle $\xymatrix{\bul{A}\ar[r]^1&\bul{A}\ar[r]&0\ar[r]&A[1]}$
is considered distinguished by hypothesis in a derived category. In the
light of the interpretation of a shifted complex as an anti-brane,
this triangle is taken to represent the annihilation of a
brane-anti-brane pair. 
Triangles in a derived category lead to what is known as the \emph{fearful
symmetry}. In the short exact sequence 
$\xymatrix{0\ar[r]& \bul{A}\ar[r]& \bul{B} \ar[r]& \bul{C}\ar[r]& 0}$ 
we can think of $\bul{B}$ as being ``made up of" $\bul{A}$ and
$\bul{C}$, with $\bul{C}$ over $\bul{A}$.  But in the derived category, 
corresponding to a triangle
\begin{equation}
\label{trg1}
\xymatrix{\bul{A}\ar[r]&\bul{B}\ar[r]&\bul{C}\ar[r]&\bul{A}[1]},
\end{equation}
there exists a homotopy-equivalent triangle
\begin{equation}
\label{trg2}
\xymatrix{\bul{B}\ar[r]&\bul{C}\ar[r]&\bul{A}[1]\ar[r]&\bul{B}[1]}.
\end{equation} 
Extending the above interpretation then implies that $\bul{C}$ is
made up of objects $\bul{B}$ and $\bul{A}[1]$, with $\bul{A}[1]$ over
$\bul{B}$. 
In particular, the notion
of \emph{subobjects} ceases to exist. Physically, it turns out to be 
convenient to interpret $\bul{C}$ as a potential bound
state of $\bul{A}$ and $\bul{B}$, corresponding to the triangle
\eq{trg1}. From the triangle \eq{trg2}, however, 
$\bul{A}$, shifted by unit degree,
appears as a bound state of $\bul{B}$ and $\bul{C}$.
Similarly, even $\bul{B}$ can be thought of as a potential bound state
of $\bul{A}$ and $\bul{C}[-1]$, by reading the triangle \eq{trg1}
toward left.
This upsets any notion of order in the derived category. 
Our goal, on the other hand, is to study \emph{stable} BPS branes. 
Any notion of stability calls for an order --- between an
object and its components. At this point let us only remark that
while such an order is not defined for the
objects in a derived category, the Abelian category does allow for an
order. Moreover, it is apparent from the discussion above that the same
triangulated category can be represented as the derived category of
different Abelian categories. These will be relevant in ascertaining
stability of branes, while still defining them to be objects in a
derived category.

Finally, let us discuss the \emph{octahedral axiom}, which can be used
in realizing an isomorphism between mapping cones of two composed
maps. Given two distinguished triangles 
\begin{equation} 
\label{octa1}
\begin{split}
\xymatrix{\bul{A}\ar[r]^f&\bul{B}\ar[r]^g&\bul{C}\ar[r]^h&\bul{A}[1]},\\
\xymatrix{\bul{B}\ar[r]^{\beta}&\bul{E}\ar[r]^{\gamma}&\bul{F}\ar[r]^{
\delta}&\bul{B}[1]},
\end{split}
\end{equation} 
there exists a complex $\bul{M}$ and two distinguished triangles,
\begin{equation}
\begin{split}
\xymatrix{\bul{A}\ar[r]^{\beta\circ f}&\bul{E}\ar[r]^{\tau}&
\bul{M}\ar[r]^{\omega}&\bul{A}[1]},\\
\xymatrix{\bul{C}\ar[r]^{\sigma}&\bul{M}\ar[r]^{\nu}&\bul{F}
\ar[r]^{g[1]\circ\delta}&\bul{C}[1]}, 
\end{split}
\end{equation}
such that the diagram 
\begin{equation}
\label{octa2}
\begin{split}
\xymatrix{
\bul{A}\ar[r]^f\ar[d]_{1}&\bul{B}\ar[r]^g\ar[d]_{\beta}
&\bul{C}\ar[r]^h\ar[d]_{\sigma}&\bul{A}[1]\ar[d]_{1}\\
\bul{A}\ar[r]^{\beta\circ f}\ar[d]_f&\bul{E}\ar[r]^{\tau}\ar[d]_{1}
& \bul{M}\ar[r]^{\omega}\ar[d]_{\nu}&\bul{A}[1]\ar[d]_{f[1]}\\
\bul{B}\ar[r]^{\beta}\ar[d]_g&\bul{E}\ar[r]^{\gamma}\ar[d]_{\tau}
&\bul{F}\ar[r]^{\delta}\ar[d]_{1}&\bul{B}[1]\ar[d]_{g[1]}\\
\bul{C}\ar[r]^{\sigma}&\bul{M}\ar[r]^{\nu}&\bul{F}\ar[r]^{g[1]
\circ\delta}&\bul{C}[1] 
}
\end{split}
\end{equation}
commutes. Even though it does not appear obvious, the above diagram
can indeed be cast into an octahedral form. However, drawing it in
that fashion does not make reading of the diagram any easier.
We shall discuss an application of this in \S\ref{centch}.
\subsection{A glimpse of the heart}
The formalism of t-structures is a means to vivisect a triangulated
category and identify the 
various Abelian subcategories in it. Let us recall some related
definitions.

For an Abelian category $\A$ and its derived category $D(\A)$, let
$\dg{n}$ denote the full subcategory, that is, one whose morphisms
coincide with the morphisms of 
the parent category, of $D(\A)$ formed by complexes
$A^{\bullet}$ with cohomology only beyond $n$, that is,
$H^i(A^{\bullet})=0$ for $i<n$.
Similarly, let $\dl{n}$ denote the full subcategory of $D(\A)$ of
complexes $A^{\bullet}$ with cohomology only below $n$, that is
$H^i(A^{\bullet})=0$ for $i>n$. A t-structure on a triangulated
category is a pair $(\dl{0},\dg{0})$ of strictly full subcategories,
satisfying the following conditions.
\begin{enumerate}
\item $\dl{0}\subset\dl{1}$ and $\dg{1}\subset\dg{0}$,
\item $\Hom^0(\dl{0}, \dg{1})=0$,
\item For each object $K$ of the triangulated category, there exists a
distinguished triangle 
\[ \xymatrix@C5mm{
& K_{\geq 1}\ar@{-->}[dr] & \\
K \ar[ur] & & K_{\leq 0}\ar[ll]
}\]
\end{enumerate}
Further, a t-structure is called \emph{bounded} if 
each $K$ in the
triangulated category is contained in $\dg{m}\cap\dl{n}$ 
for some integers $m$ and $n$.
The intersection $\dg{0}\cap\dl{0}$
coincides with $\A$, now called the \emph{heart} of the t-structure.
A bounded t-structure is completely described by its heart. 
We use bounded t-structures only and describe them using the heart. 
Physically, the heart of a t-structure provides the basic 
brane configurations with which we can construct the others.
Let us now turn to discussing the stability criterion for objects in a
derived category.
\subsection{Stable branes}
The definition of a stability criterion in a triangulated category, of
which a derived category is an example, owes its origin largely to the physics
of BPS branes\cite{douglas3}. Mathematically, it is a generalization of the
$\mu$-stability condition in the Abelian category of coherent sheaves
on a non-singular projective curve, defined using a Harder-Narasimhan
filtration\cite{brid1,brid2,brid3,macri}. 
As stressed above, a stability criterion requires sorting the objects,
which in turn calls for the notion of an index. While a triangulated
category obfuscates such a notion, an Abelian category is amenable to
it. Thus, to impose an order on a triangulated category one first
identifies an Abelian category within a triangulated category
using t-structures and then orders the objects in the Abelian
category. The latter is achieved through defining
a \emph{centered slope function} on an Abelian category $\A$ as a
group homomorphism $Z: K(\A)\rt\C$, from the Grothendieck group
$K(\A)$ of $\A$, such that for all non-zero objects $E$ of $\A$ the
complex number $Z(E)$ lies in the strict upper-half plane --- the
upper half of the complex plane sans the real axis. The \emph{phase}
of a non-zero object $E$ of $\A$ is defined in terms of the slope
function as 
\begin{equation}
\vphi (E)=-\frac{1}{\pi}\arg Z(E)\in [-1,0).
\end{equation}
The non-zero objects of the Abelian category $\A$ can be ordered by
their phases, $\vphi$. Indeed, a non-zero object $E$ of $\A$ is said
to be \emph{semistable} if $\vphi(A)\leq\vphi(E)$ for every non-zero
subobject $A$ of $E$. A slope function $Z$ defined on $\A$ is said to
have the \emph{Harder-Narasimhan property} if every non-zero object of
$\A$ has a Harder-Narasimhan filtration. That is, for each object $E$
of $\A$, there is a finite chain of subobjects 
\begin{equation*}
0=E_0\subset E_1\subset\cdots\subset E_{n-1}\subset E_n=E,
\end{equation*}
such that the quotients $F_i = E_i/E_{i-1}$ are semistable objects of
$\A$ with 
\begin{equation}
\label{hnvp}
\vphi(F_1) > \vphi(F_2) >\cdots > \vphi(F_n).
\end{equation}
We have remarked above that the heart of the t-structure is an Abelian
category. Specifying a stability condition on a triangulated category
is equivalent to giving a bounded t-structure on it along with a
centered slope function on the heart of the t-structure with the
Harder-Narasimhan property\cite{brid1}.
This is called the $\Pi$-stability, $\Pi$ being the
traditional symbol for periods of  homology cycles.
\subsection{The central charge}\label{centch}
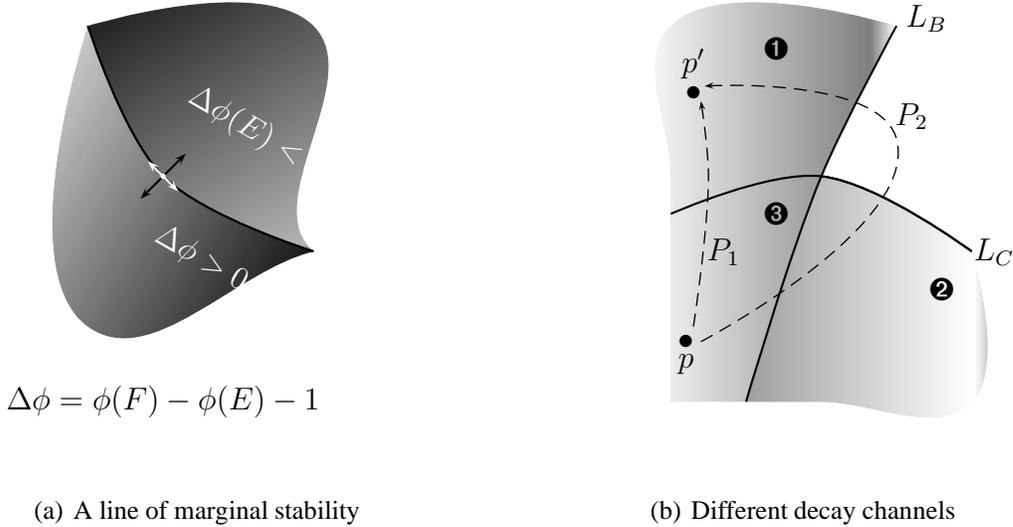
\begin{figure}[h]
\begin{center}
\subfigure[A line of marginal stability]{%
\label{fig:msl}
\begin{pspicture}(-2.1,-4)(3,2.4)
\pscustom{
\pscurve(-1,2)(0,0)(2,-1)
\gsave
\pscurve(2,-1)(2,2)(-1,2)
\fill[fillstyle=gradient,gradbegin=black,gradend=lightgray,gradangle=30]
\grestore
}
\rput[br]{-30}(2,-.1){$\white\Delta\phi(E)<0 $}
\pscustom{
\pscurve(-1,2)(0,0)(2,-1)
\gsave
\pscurve(2,-1)(-1,-2)(-1,2)
\fill[fillstyle=gradient,gradbegin=white,gradend=black,gradangle=60]
\grestore
}
\rput[br]{-30}(1,-1.6){$\white\Delta\phi>0 $}
\rput(0,-3){$\Delta\phi=\phi(F)-\phi(E) - 1$}
\psline{->}(0,0)(.3,.3)
\psline{->}(0,0)(-.3,-.3)
\psline[linecolor=white]{->}(0,0)(.2,-.2)
\psline[linecolor=white]{->}(0,0)(-.2,.2)
\psset{linecolor=white}
\qdisk(0,0){1pt}
\end{pspicture}
}%
\hskip 1in
\subfigure[Different decay channels]{
\label{fig:octo}
\begin{pspicture}(-3,-4)(2.5,3)
\pscustom{
\pscurve(-2,-.5)(0,0)(2,-1)
\gsave
\pscurve(2,-3)(0,-3)(-1,-3)
\fill[fillstyle=gradient,gradbegin=gray,gradend=white,gradangle=90]
\grestore
}
\pscustom{
\pscurve(-1,-3)(0,0)(1,2)
\gsave
\pscurve(-2,2)(-2,0)(-2,-3)
\fill[fillstyle=gradient,gradbegin=white,gradend=gray,gradangle=90]
\grestore
}
\pscurve(-2,-.5)(0,0)(2,-1)
\pscurve[linestyle=dashed,linewidth=.5pt]{->}(-1.7,-2)(-1.5,0)(-1.6,1)
\pscurve[linestyle=dashed,linewidth=.5pt]{->}(-1.6,-2.2)(1,.5)(-1.55,1.2)
\rput(-1.8,-2.2){$\bullet$}
\rput(-1.8,-2.5){$p$}
\rput(-1.7,1.1){$\bullet$}
\rput(-1.7,1.5){$p'$}
\rput(2.3,-1){$L_C$}
\rput(1.4,2.1){$L_B$}
\rput(-1.3,-1){$P_1$}
\rput(1.2,.8){$P_2$}
\rput(-.6,1.7){\ding{202}}
\rput(1.6,-1.5){\ding{203}}
\rput(-.6,-.5){\ding{204}}
\end{pspicture}
}
\end{center}
\caption{Marginal stability}
\end{figure}

The portrayal of stable D-branes as objects in the derived category of
coherent sheaves on the target space thus necessitates specifying a 
centered slope
function with the Harder-Narasimhan property. For B-type branes such a
slope function is furnished by the central charge of the 
superconformal algebra of BPS branes. Indeed, it was this central charge that  
motivated the mathematical
definition of the slope function. We denote the
central charge of a brane by $Z$ too. 
The phase $\phi$ of $Z$, called the \emph{grade}, defines a centered
slope function with the Harder-Narasimhan property determining 
the stability of branes. Let us consider the type--IIA theory compactified 
on a Calabi-Yau manifold, $\M$.  
Let $E$ and $F$ be two D-branes in the derived category 
$D^b(\coh\M)$ and let $C = M(f: E\rt F)$ be the mapping cone mentioned
earlier, which corresponds to a marginal deformation of the branes $E$
and $F$, a potential bound state. If it is marginally stable, then
we have, $Z(C) = Z(F)-Z(E)$, implying 
\begin{equation}
\label{ms}
\phi(C) = \phi(F) = \phi(E)+1.
\end{equation}
The grade $\phi$ depends on the K\"ahler parameter of $\M$ and thus
given two branes $E$ and $F$ in the derived category, \eq{ms}
determines the set of points in the K\"ahler moduli space where the
bound state $C$ is marginally stable
\cite{sen2,mrd:ictp,douglas1,Sharpe:1999qz,douglas2,douglas3,dfr1,dfr2}. 
This set is called the
\emph{line of marginal stability}. Given a point on this line, three
cases arise as we move off this point in the K\"ahler moduli space,
namely,
$\Delta\phi\equiv\phi(F) - \phi(E)-1 \lesseqqgtr  0 $. If the expression is negative,
then an open string joining the branes becomes tachyonic and 
$C$ becomes a bound state. If $\Delta\phi$ is positive, on the other hand, then $C$
is unstable against decay to $E$ and $F$. If the expression continues
to be equal to zero, then the bound state remains settled into a stasis on 
the line of marginal stability.
This is illustrated in the \figref{fig:msl}.
Thus, an analysis of $\Pi$-stability involves two steps. First, the
marginal stability line is to be obtained. Then the stability of
branes is to be checked as a second step as one goes off this line. 
If it all seems copacetic so far, let us note that
complications arise in ascertaining the stability of branes
from stability issues of the components. For example, even if
the expression $\phi(F) - \phi(E)-1 $ is negative, it is not possible to aver
that $C$ is stable. Indeed, it is not improbable that  
the derived category contains another triangle with $C$ at a vertex.
$C$ may then be a bound state of certain other branes, $E'\rt F'$. 
In other words, while $C$ is stable with respect to decaying into $E$ 
and $F$, it may have other decay channels. If, moreover, the expression 
is positive, we can not conclude with certitude
that $C$ is unstable, since the components $E$ and
$F$ may themselves be unstable and then the bound state may be 
perdurable, being energetically favored. A quandary as such can be settled 
if we have means to compare two decay channels, or, in other words, we can 
decide whether two distinguished
triangles are isomorphic or not. As discussed before, this can be done
using the octahedral axiom. We now proceed to discuss the issue in
some detail
as an illustration of the intricate internal consistency of the definitions
\cite{asp-doug}.

Let us consider an object $C$ which is known to be stable at a given
point in the moduli space $p$, marked in \figref{fig:octo}. Let a decay
channel of $C$ be $C\rt A + B$ across the line $L_C$ with
$\phi(B)=\phi(A)+1$ on the line. Thus, if $A$ and $B$ are both 
stable, then $L_C$
is a line of marginal stability for this decay, with $C$ stable in the
region \ding{203}$\cup$\ding{204}. Similarly, let us suppose
that there is another possible decay $B\rt E+F$ across the line $L_B$,
$B$ being stable in the region \ding{202}$\cup$\ding{204}.
Let us now consider two different paths $P_1$ and $P_2$ from $p$ in
region \ding{204} to a point $p'$ in region \ding{202}. Tracing 
the path $P_1$, $C$ decays across $L_C$.
But along the path $P_2$ which is homotopic to $P_1$, 
on the other hand, $L_B$ is crossed
before $L_C$; $B$ has decayed before arriving at $L_C$. This
challenges our hypothesis of stability of $C$ in the region \ding{203}.
Let us now appeal to the octahedral axiom, \eq{octa2}. 
Given the two bound states $C=A+B$ and $B=E+F$, 
we have two triangles $\xymatrix{A\ar[r]&B\ar[r]&C\ar[r]&A[1]}$ and 
$\xymatrix{E\ar[r]&F\ar[r]&B\ar[r]&E[1]}$, respectively.
Let us rewrite the latter one in the homotopy equivalent form 
$\xymatrix{B\ar[r]&E[1]\ar[r]&F[1]\ar[r]&B[1]}$, in order to bring the
triangles to the form \eq{octa1}.
Let us define
$\phi_0 = \phi(A)$ and evaluate the grades of other objects using
\eq{octa2}, with $E$ and $F$ replaced with $E[1]$ and $F[1]$,
respectively and assuming that we are on the line $L_C$ in the region 
\ding{202}$\cup$\ding{204}.
We write the grades in a tabular form corresponding to
the objects in the diagram \eq{octa2}:
\begin{equation}
\begin{split}
\xymatrix{
{A}\ar[r]\ar[d]&{B}\ar[r]\ar[d]
&{C}\ar[r]\ar[d]&{A}[1]\ar[d]\\
{A}\ar[r]\ar[d]&E[1]\ar[r]\ar[d]
&{M}\ar[r]\ar[d]&{A}[1]\ar[d]\\
B\ar[r]\ar[d]&{E}[1]\ar[r]\ar[d]
&{F}[1]\ar[r]\ar[d]&{B}[1]\ar[d]\\
{C}\ar[r]&{M}\ar[r]&{F}[1]\ar[r]&{C}[1] 
}
\quad
\xymatrix@C.2cm{
\phi_0 & \phi_0+1 & \phi_0+1&\phi_0+1\\
\phi_0 & \phi_0+1 & \phi_0+1-\varepsilon & \phi_0+1\\
\phi_0+1 & \phi_0+1-\varepsilon'& \phi_0+2+\varepsilon''& \phi_0+2\\
\phi_0+1 & \phi_0+1-\varepsilon &\phi_0+2&\phi_0+2
}
\end{split} 
\end{equation} 
where $\varepsilon$, $\varepsilon'$ and $\varepsilon''$ are positive
numbers. Let us now trace $L_C$ from left to right, by changing the
K\"ahler modulus. $C$ continues to be a bound state, while
destabilizing $B$ in the process to decay into $E+F$. That is, we have 
\begin{equation} 
\phi(F) - \phi(E) -1 >0.
\end{equation} 
Since we have not altered $\phi(A)$ and hence $\phi(B)$, we derive
from the table that in this process $\phi(F)$ increases while
$\phi(E)$ and $\phi(M)$ decrease. Then by the second row of the
commutative diagram,
\begin{equation} 
\phi(E[1]) - \phi(A) = 1 -\varepsilon' < 1,
\end{equation} 
ergo $M$ is now stable. Thus, while $B$ decays across $L_B$, the 
bound state $C$
continues to be a stable state, a bound state of objects different
from $A$ and $B$ and decays only across $L_C$. 
We thus learn that
while the decay of a brane into two
daughters is studied using a distinguished triangle, the decay of a
brane into three needs invoking the octahedral axiom. In order to
consider multiple decays we need to consider a set of distinguished
triangles corresponding to a brane $E$\cite{aspin1,brid1}. 
If $E$ can be written as 
\begin{equation}
\label{HN}
\begin{split}
\xymatrix@C3mm{%
0=E_0 \ar[rr]&&E_1\ar[rr]\ar[dl]&&E_2\ar[rr]\ar[dl]
&&\cdots\ar[rr]\ar[dl]&&E_{n-1}\ar[rr]\ar[dl]&&E_n=E\ar[dl]\\
& A_1\ar[ul]^{[1]} && A_2\ar[ul]^{[1]} && A_3\ar[ul]^{[1]}  
&& A_{n-1}\ar[ul]^{[1]}  && A_n\ar[ul]^{[1]} 
}%
\end{split}
\end{equation}
then it may decay into $A_1, A_2,\cdots A_n$, if 
\begin{equation}\label{HNphi}
\phi(A_1) > \phi(A_2) > \cdots > \phi(A_n).
\end{equation}
Let us remark that this is where the Harder-Narasimhan property of
the slope function \eq{hnvp} is put to use. However, 
as we have discussed above,  for the decay to
take place, the decay products have to be stable. It is deemed
that at each point of the complexified K\"ahler moduli space of $\M$
there exists a set of stable D-branes so that the object in
$D^b(\coh\M)$ corresponding to any given brane can be expressed as
\eq{HN} for a certain integer
$n$ and a set of stable objects $\{A_n\}$ satisfying
\eq{HNphi}. The stable objects are to be chosen
from the heart of a t-structure, which by virtue of being an Abelian 
category is
amenable to the notion of an order. Thus, to every point of the K\"ahler
moduli space should be assigned a t-structure in whose heart the
stable objects $A_n$ take residence. This requires mapping out the
t-structures of $D^b(\coh\M)$ on the K\"ahler moduli space of $\M$. 
This is a formidable task in general and not much
is known about the t-structures for Calabi-Yau manifolds.
However, the t-structures on $D^b(\coh\cp{1})$ have been classified.
In the next section we shall consider an assignment of t-structures
to the points in the K\"ahler moduli space of $\cp{1}$, which arises
from a geometrically engineered Calabi-Yau.

Before this, let us briefly discuss some features of the grade which
will be useful in the calculations. As remarked before, at each point
in the K\"ahler moduli space, a set of stable \emph{basic} brane is 
deemed to exist and any brane in the spectrum is to be made up of 
these basic ones. The stability of a
brane is ascertained with respect to decaying into these basic branes.
But this surmises the stability of the
basic branes a priori. This leads to a circularity in
the criterion of stability that we have discussed so far. The
resolution is to assume that there is some other way of determining
the class of stable objects at a certain given point in the moduli
space, consistent with $\Pi$-stability, 
and then apply the definition in terms of the grade 
$\phi$ as other points of the moduli space is traversed. One
convenient choice for the other criterion is the $\mu$-stability
of sheaves in some region in the moduli space. But this necessitates
the identification of an Abelian category of sheaves in the region.
We need to know the different t-structures of the derived
category realized in various regions in the K\"ahler moduli space. 
To the \emph{large volume} limit of $\M$, which is a region in the
K\"ahler moduli space, where geometric notions are valid, 
we then ascribe the Abelian category $\coh\M$ and 
employ the orthodox $\mu$-stability of sheaves to
determine if a brane is stable in this region. Once this 
calibration is worked out,
we can move along paths penetrating the deep  
interior of the K\"ahler moduli space and study stability of branes.

To be more concrete, let us consider type--IIA theory on
$\R^{1,3}\times\M$, where $\M$ is a Calabi-Yau
manifold\cite{tj,suresh,dfr1,dfr2,aspin1,mrd:ictp}.  The
topological D-branes in the corresponding B-model are objects of the
bounded derived category of coherent sheaves on $\M$, 
$D^b(\coh{\M})$. For an object $E$ of $\coh{\M}$ the central charge is
given by 
\begin{equation}
Z(E) = -\int\limits_{\M}e^{-tK}\ch{E}\sqrt{\td{\M}} + \cdots
\end{equation}
in the large volume limit,
where $K$ denotes the generator of the K\"ahler cone, $t$ the
K\"ahler parameter, complexified with the anti-symmetric field $B$
which appeared in the Dirac-Born-Infeld action \eq{DBI}, 
$\ch{E}$ denotes the Chern character of $E$
and $\td{\M}$ denotes the Todd class of $\M$ and
the integration is over $\M$. The grade is then given by 
\begin{equation}
\phi =  -\frac{1}{\pi} \operatorname{Im}\log Z(E).
\end{equation}
In order to study the stability of branes in the K\"ahler moduli space
of $\M$, then, it turns out to be convenient to fix a base-point in the
large volume limit where the basic branes are taken to be the ones
residing in $\coh\M$, which is the heart of the canonical t-structure
of $D^b(\coh\M)$, as mentioned before. It is with respect to this
base-point that we assign grades to the branes and t-structures 
at other points in the moduli space.
\section{A heart for a stable brane}
Let us now discuss an application of the
various aspects of $\Pi$-stability discussed so far with a
simple example. This serves
to illustrate the usefulness of the intricate machinery 
in a simple physical problem. In particular, 
we describe a limit of a string theory compactification where
the BPS states naturally arise as objects of derived category of
coherent sheaves on the projective curve $\cp{1}$. We also work 
out the atlas of
t-structures over the K\"ahler moduli space of $\cp{1}$.
\subsection{Degenerating the target}
Let us consider a Calabi-Yau threefold
which is a $\mathbf{K3}$ fibration on $\cp{1}$. 
The compactified theory
has a gauge sector with gauge fields corresponding to the
open strings connecting D-branes wrapped on various compact homology
cycles of the Calabi-Yau.
Choosing the volume of the projective curve to be large and that of the
$\mathbf{K3}$ appropriate, we obtain a non-Abelian gauge theory.

More specifically, let us consider the space $\M$ as
the degree 8 hypersurface in
the resolution of the weighted projective space $\cp{4}[2,2,2,1,1]$.
In order to study phenomena in the moduli space of the 
complexified K\"ahler two-form of $\M$ including non-perturbative
influences we have recourse to Mirror symmetry and 
consider the mirror dual $\W$ of $\M$ which is given by
a $\Z_4\times\Z_4\times\Z_4$ quotient of the same hypersurface and is
described by the equation
\begin{equation}
a_0z_1z_2z_3z_4z_5 + a_1z_1^4 + a_2z_2^4 + 
a_3z_3^4 + a_4z_4^8 + a_5z_5^8 + a_6 z_4^4z_5^4 = 0,
\end{equation}
where $z_1,\cdots, z_5$ denote the coordinates of the five-dimensional
affine complex space $\C^5$ and
the parameters $a_i$ represent complex deformations of the polynomial
thus corresponding to deformation of complex structure.
The algebraic coordinates of the complex structure moduli
space are 
\begin{equation}
x = \frac{a_1a_2a_3a_6}{a_0^4}
\quad
y = \frac{a_4 a_5}{a_0^2},
\end{equation}
obtained by rescaling the affine coordinates $\{z_i\}$.
The Mirror map relates these coordinates to the
complexified K\"ahler moduli. The two K\"ahler moduli corresponding to
$x$ and $y$, denoted $(B+iJ)_x$ and $(B+iJ)$, represent, respectively,
the sizes of $\mathbf{K3}$ and $\cp{1}$. These are obtained as
solutions to the Picard-Fuchs equations for the periods and are given
by 
\begin{equation}
(B+iJ)_x = \frac{1}{2\pi i}\log{x} + {\mathcal O}(x,y),\quad
(B+iJ)_y = \frac{1}{2\pi i}\log{y} + {\mathcal O}(x,y),
\end{equation}
where ${\mathcal O}(x,y)$ represents linear and higher order terms in
$x$ and $y$.
The mirror $\W$ is an algebraic variety which is singular along 
a \emph{discriminant locus}, $\triangle = \triangle_0\triangle_1^3$, 
with the \emph{primary} component,
$\triangle_0$, given by 
\begin{equation}
(1-2^8 x)^2 - 2^{18} x^2 y = 0,
\end{equation}
and $\triangle_1$ given by $1-4y=0$.
In the limit in which the volume of the base $\cp{1}$ is large, that
is $y\rt 0$, the gauge symmetry of the configuration enhances to a
non-Abelian one, namely, $SU(2)$. The non-perturbative nature of this
enhancement dictates restricting our considerations to
the discriminant locus, thereby forcing a choice of $x=1/2^8$ too.
By the double scaling limit of $(x,y)\rt(2^{-8},0)$ and setting 
$u = \frac{ 1 - 2^8 x }{2\sqrt{y}}$ we can consider the K\"ahler
moduli space in the neighborhood of the point of enhanced symmetry.
This procedure is called \emph{geometric engineering}\cite{vafa}.
The moduli space of interest is now the $u$-plane, intersecting 
the principal component $\triangle_0$ at $u=\pm 1 $. 
This can be identified as the $u$-plane of the 
Seiberg-Witten $SU(2)$ theory. The supersymmetric states in this theory
arise from the branes in $\M$ which become massless at 
$(x,y)=(2^{-8},0)$.
We envisage these branes as being objects in the derived category of coherent
sheaves on $\cp{1}$\cite{AHK}. This construction provides a simple
physical arena for the derived category picture of branes to be 
realized. Indeed, for sufficiently large volumes 
all the objects in $\dercat{\coh{\cp{1}}}$
can be generated by the structure sheaf $\sheaf{O}_{\cp{1}}$ on
the projective curve, physically
corresponding to a D4-brane wrapped on the $\mathbf{K3}$ and 
the sky-scrapper sheaf $\O_x$ which corresponds to a D6-brane 
wrapped on the whole of $\M$, where $x$ is a point in $\cp{1}$.

It thus suffices to consider $\Pi$-stability on the derived category
of coherent sheaves on $\cp{1}$. The stability
condition requires defining a centered
slope function which is a map from the K-group of $\cp{1}$
to $\C$. The K-group of $\cp{1}$ is generated by $H^0(\cp{1})$ 
and $H^2(\cp{1})$. We can take the two generators of 
$\dercat{\coh{\cp{1}}}$ namely $\O_{\cp{1}}$ and $\O_x$ to be
the generators of the K-group. The Chern characters 
of these sheaves span $H^0(\cp{1})$ 
and $H^2(\cp{1})$respectively. Therefore, to obtain the slope 
function of an object we require two basic slopes, determined by the
phases of the central charges, 
\begin{equation}
a = Z(\pi^{\star}\O_x),\quad
a_D = Z(\pi^{\star}\O_c),
\end{equation}
where $a$ and $a_D$ are given in terms of the periods of  homology
cycles. Here $\pi^{\star}$ denotes the pull-back of the map
$\pi:\M\rt\cp{1}$, signifying that the branes in $\M$ are obtained
from the sheaves on $\cp{1}$.
The periods are obtained as solutions to the Picard-Fuchs equation associated
with the Calabi-Yau $\M$, which in the double-scaling limit becomes 
\begin{equation}
z(1-z) \frac{\partial^2 \Phi}{\partial z^2} - \frac{1}{4}\Phi = 0,
\end{equation}
where we defined $z = (1/2)(u-1)$. The periods are obtained as
the two solutions to this equation as
\begin{gather}
a(u) = \sqrt{2(u+1)} {}_2F_1(-\frac{1}{2} , \frac{1}{2}, 1; \frac{2}{u+1}),
\\
a_D(u) = -\frac{u-1}{2i} {}_2F_1(\frac{1}{2} , \frac{1}{2}, 2;
\frac{2}{1-u}),
\end{gather}
where the Hypergeometric functions are defined on the 
$u$-plane with  branch-cuts from $(-1,\infty)$ for $a$ and
$(1,\infty)$ for $a_D$.
Slope functions for an arbitrary sheaf on $\cp{1}$ 
can be obtained from these expressions for the periods. For
example, the central charge of $\pi^{\star}\O_{\cp{1}}(n)$ is given by 
\begin{equation}
Z(\O(n)) = a_D(u) + n a(u),
\end{equation}
and its phase determines the slope function at any point $u$.

The simplicity of this example owes it origin mainly to the fact
that there is but a single curve of marginal stability, an ellipse,
homotopic to the circle $|u|=1$ and passing through $u=\pm 1$, as
shown by the dotted line in \figref{fig:sw}. 
The stable states outside the ellipse
are $\pi^{\star}\O_{\cp{1}}(n)$, with $n\in\Z$ and 
$\pi^{\star}\O_x$, with $x\in\cp{1}$.
Inside the circle the stable states are only $\pi^{\star}\O_{\cp{1}}$ and
either of $\pi^{\star}\O_{\cp{1}}(1)$ and
$\pi^{\star}\O_{\cp{1}}(-1)$,
depending on the direction of entry into the interior. 
These reproduce the the totality of states in the 
dyon spectrum of the Seiberg-Wittten theory both
in the weak and the strong coupling regimes, which are identified as
the regions outside and inside of the ellipse, respectively. The string
junctions and spiral strings on the moduli space have also been
identified among these stable objects\cite{karp,mmr}.
\subsection{Hearts with perverse sheaves}
The t-structures bring out 
the Abelian categories within a derived category.
T-structures on the derived category 
$\dercat{\coh\cp{1}}$ have been classified \cite{rudakov} upto
autoequivalenes of the derived category, forming a group
$\mathfrak{Aut}(\dercat{\coh{\cp{1}}})$.
As a preparation to constructing the atlas of 
t-structures on the moduli space let us
first take stock of the bounded t-structures on
$\dercat{\coh{\cp{1}}}$.
There are two classes of t-structures on $\dercat{\coh\cp{1}}$,
namely, \emph{standard} and \emph{exceptional}. We
begin with the classification of the standard t-structures.

The simplest t-structure on $\dercat{\coh\cp{1}}$ is the tautological
one with $\coh\cp{1}$ as its heart, given by
\begin{equation}
\tA^{\leq 0} = \langle \coh{\cp{1}}[j], j \geq 0\rangle,
\tA^{\geq 0} = \langle \coh{\cp{1}}[j], j \leq 0\rangle.
\end{equation}
The heart is $\tA^{\heart} = \coh{\cp{1}}$ which contains only branes
and no antibranes. This is the t-structure that we posit to be
realized in the large volume limit.

Other standard t-structures are obtained from $\coh\cp{1}$ by constructing
cotilting torsion pairs. We shall not delve into the details of the 
construction.
For our purposes it suffices to recall that whenever an Abelian category $\A$
can be split into a torsion pair $(\A_1,\A_0)$ of full subcategories
satisfying certain criteria \cite{rudakov,manin}, 
we can obtain a t-structure from it as
\begin{gather*}
\phantom{A}^p\dercat{\A}^{\leq 0} = 
\{ A\in\dercat{\A}^{\leq 0} | H^0(A)\in\A_1\},\\
\phantom{A}^p\dercat{\A}^{\geq 0} = 
\{ A\in\dercat{\A}^{\geq -1} | H^{-1}(A)\in\A_0\},
\end{gather*}
whose heart is the category of $p$-perverse sheaves on
$\cp{1}$\cite{manin}.
We can thus start from the Abelian category $\coh{\cp{1}}$ 
obtained as the heart
of the tautological t-structure and obtain others with perverse
sheaves through cotilting. The first torsion pair that 
is constructed in this vein is
\begin{equation}
\tB_0 = \langle \O(n), n<0 \rangle,\quad
\tB_1 = \langle \O(n), n \geq 0 ; \O_x , x \in \cp{1} \rangle.
\end{equation}
The associated t-structure, then, is given by
\begin{equation}
\tB^{\leq 0} = \langle \O(n)[i], n \geq 0 , i \geq 0 ;
 \O_x[i] , x \in \cp{1}, i \geq 0 ; 
\O(n)[j] , n < 0 , j \geq 1 \rangle ,
\end{equation}
where we have recorded only one half of the t-structure
since a t-structure is unambiguously specified by a
moiety. The heart of the t-structure is
\begin{equation}
\tB^{\heart} = \langle \O(n)[0] , n \geq 0 ; \O_x[0] , x 
\in \cp{1} ; \O(n)[1] , n < 0 \rangle
\end{equation}
The heart is different from that of the tautological t-structure.
In particular, let us remark that the heart $\tB^{\heart}$ contains 
objects which would be
interpreted as antibranes in the parlance of the large volume regime.

There are two other cotilting torsion pairs that completes
the list of standard bounded
t-structures on $\dercat{\coh\cp{1}}$. One of these is
\begin{equation}
\tC_0 = \langle \O(n), n \in \Z \rangle,\quad
\tC_1 = \langle \O_x , x \in \cp{1} \rangle,
\end{equation}
giving rise to the t-structure
\begin{equation}
\tC^{\leq 0} = \langle \O(n)[i], n \in \Z , i \geq 0 ;
 \O_x[i] , x \in \cp{1}, i \geq 0 ;\rangle ,
\end{equation}
with the heart
\begin{equation}
\tC^{\heart} = \langle \O(n)[1] , n \in \Z ; 
\O_x[0] , x \in \cp{1} \rangle.
\end{equation}
The other torsion requires an arbitrary nonempty subset $P \subset \cp{1}$
\cite{rudakov}.
This t-structure is not realized on the moduli space. 
So we refrain from listing it.

There are two kinds of exceptional t-structures: 
\emph{bounded} and \emph{unbounded}.
As mentioned earlier,  we are interested only in the 
bounded t-structures of which
there are but two in the derived category  $\dercat{\coh{\cp{1}}}$.
These depend on an integer $k \in \Z$, and are given by 
\begin{gather}
\tE^{\leq 0} = \langle \O[i] , i \geq k ; \O(1)[j] , j \geq -2,\\
\tF^{\leq 0} = \langle \O[i] , i \geq k ; \O(1)[j] , j \geq -1,
\end{gather}
with hearts 
\begin{gather}
\tE^{\heart} = \langle \O(k)[0] , \O(1)[-2]\rangle,\\
\tF^{\heart} = \langle \O(k)[0] , \O(1)[-1]\rangle ,
\end{gather}
respectively.
\begin{figure}[h]
\begin{center}
\subfigure[Marginal stability line and stable branes]{%
\label{fig:sw}
\begin{pspicture}(-2,-2.4)(4,2)
\psset{unit=1.5cm}
\psellipse[linestyle=dotted](0,0)(1.2,1)
\psaxes[ticks=none,labels=none]{<->}(0,0)(-1.7,-1.5)(1.7,1.5)
\rput(-1.5,.2){$-1$}
\rput(1.5,.2){$1$}
\rput(1.5,1){
$\scriptstyle\pi^{\star}\O_{\cp{1}}(n),\,\,\,
\pi^{\star}\O_x$}
\rput(0.1,-0.3){
$\scriptstyle\pi^{\star}\O_{\cp{1}}, 
\quad\pi^{\star}\O_{\cp{1}}(1)$ 
}
\end{pspicture}
}%
\hskip 1cm
\subfigure[t-structures ]{%
\begin{pspicture}(-2,-2.4)(4,3)
\psset{unit=1.5cm}
\pscustom{
\pscurve(1.2,0)(1.15,1)(.7,2)
\gsave
\pscurve[liftpen=1](2.4,2)(2,1)(1.2,0)
\fill[fillstyle=gradient,gradbegin=white,gradend=gray,gradangle=60]
\grestore
\pscurve[liftpen=2](2.4,2)(2,1)(1.2,0)
}
\pscustom{
\pscurve(1.2,0)(2,1)(2.4,2)
\gsave
\psline[liftpen=1](3,0)(1.2,0)
\fill[fillstyle=gradient,gradbegin=black,gradend=lightgray,gradangle=60]
\grestore
}
\psellipse[linestyle=none,fillstyle=solid,fillcolor=darkgray](0,0)(1.2,1)
\psline[linestyle=dashed,linecolor=white,dash= 3pt 2pt](-1.2,0)(1.2,0)
\psline[linestyle=dashed,dash= 3pt 2pt](1.2,0)(3.2,0)
\rput(2.5,.5){$\tA$}
\rput(-1.5,-.5){$\tC$}
\rput(1.5,1){$\tB$}
\rput(0,.5){$\white{\tE}$}
\rput(0,-.5){$\white{\tF}$}
\end{pspicture}
\label{fig:tst}
}
\caption{K\"ahler moduli space of $\cp{1}$}
\end{center}
\end{figure}
\subsection{Branes in the hearts}
Finally, let us discuss the variation of t-structures over the moduli
space of K\"ahler volume of $\cp{1}$. If we fix a window
in the slope function 
then the various t-structures make their appearances as we
wander about the moduli space. 
As discussed above, to obtain an atlas of t-structures in the moduli space
the slope function has to be calibrated by matching with
the $\mu$-stability criterion in the large volume limit. 
In this region  stable objects are given by the invertible sheaves or
line bundles on $\cp{1}$. Since 
the heart is expected to consist of stable objects only,
it has to contain $\O(n)$, for integers,  $n$ and the
t-structure corresponds to some standard t-structure. 
On the other hand, within the confines of the line of  marginal
stability it is only $\O$ and $\O(1)$ or $\O(-1)$, which are stable.
So the candidate t-structure in this region would be
exceptional t-structure. 
Let us mention a point about notation in the following. 
We shall consider sheaves on
$\cp{1}$ from now on. To interpret these as branes we have to 
pull-back the sheaves on
$\M$ with $\pi^{\star}$. We shall suppress both the $\cp{1}$
and the pull-back from the notation in
the following for typographical ease. 
Let us now chart out the atlas of t-structures on the K\"ahler 
moduli space of $\cp{1}$. 
\begin{dinglist}{109}
\item First, let us consider the region in the $u$-plane 
with $\mathrm{Re}~u>1$ and $\mathrm{Im}~u$ small and positive. This
includes the large radius region. Looking for objects
whose phases lies in the interval $\phi\in[-1,0)$, we find the set of objects,
$\langle \O(n) , n \in \Z ; \O_x , x\in \cp{1} \rangle$.
This set consists of branes only, with no anti-branes. We identify
this with the heart of the tautological t-structure, $\tA^{\heart}$.  
In other words, we assign the t-structure $\tA$ to this region 
of the moduli space.
This is marked in \figref{fig:tst}.
\item Let us now move away from the positive $\mathrm{Re}~u$-axis 
counterclockwise. Grades of some of the objects, namely,
$\O(-n)$, with $n > N$ for some integer $N$, will fall below $-1$, 
outside the grade 
window. While these objects are defenestrated, their shifted
cousins $\O(-n)[1]$ make an entry though the other corner of the window.
At a generic point the set of 
objects carrying grades within the window consists of
$\langle \O(n) , n\in \Z , n > -N ; \O(n)[1] , n \leq -N ; 
\O_x , x \in \cp{1} \rangle$. Apparently, these do not have a place in 
the heart of any of the t-structures listed above. But
let us recall that t-structures are classified upto 
autoequivalences of the category. One such autoequivalence
can be generated by a monodromy transformation associated
with the path followed in our journey around the large volume 
point which transforms
a sheaf by tensoring it with $\O(1)$, which results in a change in the grade. 
Transforming the objects by the monodromy $N$ times we can coerce the objects
to be in the heart of the t-structure $\tB$, as marked in \figref{fig:tst}.
These will realize the spiral strings identified earlier\cite{mmr}.
\item Continuing our journey in the $u$-plane, we find that eventually
all the $\O(n)$ are replaced by their shifted cousins
and beyond a certain line, shown in \figref{fig:tst}, it is only these cousins 
$\langle \O(n)[1] , n \in \Z ; \O_x , x \in \cp{1} \rangle$
that find a place in the heart of the t-structure $\tC$.
The t-structure $\tC$ is ascribed to the white space in
\figref{fig:tst}.
\item Continuing further around the ellipse 
as we stumble upon the positive $\mathrm{Re}~u$-axis from below, 
due the presence of branch cut introduced for defining the periods, 
the grades jump and we come back to the regime of the t-structure $\tA$.
\item The above discussion is valid outside the ellipse of
marginal stability. As long as we do not cross the
line of marginal stability, the standard t-structures reign.
The branes incarcerated within the confines of the marginal 
stability locus, on the
other hand, correspond to the exceptional t-structures.
In the region $\mathrm{Im}~u < 1$, the branes take residence in the 
heart of the t-structure $\tE$ consisting of 
$\langle\O , \O(1)[-2]\rangle$.
\item Below the $\mathrm{Re}~u$-axis the phase of $a_D(u)$ 
jumps across the branch cut. 
The basic branes now find an abode in the heart $\tF^{\heart}$ of the 
t-structure $\tF$, with the basic branes
$\langle \O , \O(1)[-1]\rangle$.
\end{dinglist}
\section{Epilogue}
We have reviewed some aspects of topological branes in the type--IIA theory 
which are
described by the B-model. The naive geometric notions turn out to be 
inadequate to
describe the D-branes precisely. 
Branes wrapped on the homology cycles of a Calabi-Yau
manifold when the type--IIA theory is compactified on it are geometrically 
portrayed as objects in
the derived category of coherent sheaves on the Calabi-Yau. We recalled several
features of a derived category before discussing the definition of 
stability of objects
and some of its subtleties. It is expected that to each point in the 
K\"ahler moduli
space of the Calabi-Yau is associated a t-structure in whose heart the basic
constituent D-branes reside. We then considered a simple example of a 
geometrically 
engineered Calabi-Yau, with only the projective curve $\cp{1}$ as 
its non-trivial part,
giving rise to the $SU(2)$ Seiberg-Witten gauge theory. The dyon
spectrum of the gauge theory is reproduced using the parlance of the 
derived category
in an elegant fashion. Finally, we associated some of the t-structures 
of the derived
category of coherent sheaves on $\cp{1}$ to different regions in the 
K\"ahler moduli space. 

\end{document}